\documentclass[a4paper,final,12pt]{article}
\usepackage{graphicx}
\begin{document}
\setcounter{page}{0}
\thispagestyle{empty}
\begin{flushright}
YITP--99--45 \\
July 1999 \\
\end{flushright}

\vspace{2cm}

\begin{center}
{\large\bf Supersymmetry and Future Colliders}
\footnote{ Talk given in International Symposium on 
Supersymmetry, Sugergravity, and Superstring(SSS99), 
Seoul, Korea, June 23(Wed)-27(Sun), 1999.
Part of this talk is based on the project in progress 
with D. Toya and T. Kobayashi, ICEPP, Tokyo University. }
\end{center}
\baselineskip=32pt

\centerline{M. M. Nojiri}

\baselineskip=22pt

\begin{center}
{\footnotesize \it YITP, Kyoto University, Kyoto, 606-8502\\
E-mail: nojiri@yukawa.kyoto-u.ac.jp}
\end{center}

\vspace{1cm}

\begin{abstract}
In this talk, I review precision SUSY study 
at LHC and TeV scale $e^+e^-$ linear colliders  (LC). 
We discuss the study of the 
3 body decay $\tilde{\chi}^0_2\rightarrow$ 
$\tilde{\chi}^0_1 l l$ or the  2 body decay $\tilde{\chi}^0_2\rightarrow 
\tilde{l}l$ at LHC. 
In the former case,  the whole $m_{ll}$ distribution 
observed at LHC
would constrain ino mixing and slepton masses. On the other hand, 
when $\tilde{l}l $ decay is open, the distribution of the  asymmetry of 
the transverse momentum of  lepton pair  $A_T= p_{T1}/p_{T2}$ peaks 
at $A_E=p_{1}/p_{2}$ at $\chi^0_2$ rest frame  
for  $m_{ll}\ll m^{\rm max}_{ll}$ samples, providing another 
model independent information. The peak position and the edge of 
the $m_{ll}$ distribution  constrain $m_{\tilde{\chi}^0_2}$,
$m_{\tilde{\chi}^0_1}$ and $m_{\tilde{l}}$. Slepton mass 
universality may be checked within a few \% in the early stage of
experiment. Finally I 
discuss the physics at TeV scale LC. The mass and couplings 
of sparticles will be measured within $O(1\%)$  error, and 
measurement of the radiative correction to the ino-slepton-lepton 
coupling will determine the first generation squark mass scale 
even in decoupling scenarios.
\end{abstract}
\vspace{2cm}
\vfill
\pagebreak

\normalsize\baselineskip=15pt
\setcounter{footnote}{0}
\renewcommand{\thefootnote}{\arabic{footnote}}

\newcommand{\tabtopsp}[1]{\vbox{\vbox to#1{}\vbox to1{}}}
\newcommand{\gsim}{\buildrel>\over{_\sim}}
\newcommand{\lsim}{\buildrel<\over{_\sim}}
\renewcommand{\thefootnote}{\fnsymbol{footnote}}
\newcommand{\tchi}{\tilde{\chi}}
\newcommand{\psla}{p\kern-.45em/}
\newcommand{\esla}{E\kern-.45em/}

\section{Introduction}
The Minimal Supersymmetric Standard Model (MSSM)  is one of 
the promising extension of 
Standard Model. If the nature  picks up  
the low energy supersymmetry(SUSY), 
MSSM  will be proven {\it for sure},
as superpartners will be copiously produced at future 
colliders such as Large Hadron Collider (LHC) at CERN or 
TeV scale $e^+e^-$ linear colliders (LC)  proposed by DESY, KEK, and SLAC. 
The symmetry also offers natural solution of the 
hierarchy problem, amazing   gauge coupling unification, 
and dark matter candidates. 

On the other hand, the MSSM suffers sever flavor changing neutral 
current (FCNC) constraints if no mass relation is imposed on  sfermion mass
parameters. Various proposals have been made 
of the mechanism to incorporate  the 
SUSY 
breaking to ``our sector'', trying to offer the  natural explanation 
of such mass relations. 
In short, it would be very surprising
if sparticles are found in any future collider--- The discovery 
is not the goal, but it is the beginning 
of a new  quest of ``the mechanism'' of SUSY breaking. 
Measurements of soft breaking masses would be an important aspects 
of the SUSY study at future colliders, because different SUSY breaking 
mechanism predict different sparticle mass patterns. 

In this talk, I will review attempts to measure soft breaking 
parameters at LHC and LC's. In section 2 and 3, 
I will concentrate on the process that $\tilde{g}$ and $\tilde{q}$ 
are produced  and decay, involving the leptonic second lightest 
neutralino decay $\tilde{\chi}^0_2 \rightarrow$ $\chi^0_1 l^+l^-$. 
The decay either proceeds through virtual exchanges of 
$Z^0$ and  $\tilde{l}$ or direct two body decays such as 
$\tilde{\chi}^0_2\rightarrow \tilde{l}l$ and $\tilde{l}\rightarrow l 
\tilde{\chi}^0_1$.
Events near the  end point of the $m_{ll}$ 
distribution of the three body decay 
play a key role to 
reconstruct the kinematics of $\tilde{g}$ and $\tilde{q}$ 
cascade decay chain, and 
minimal supergravity parameters is determined precisely.\cite{H}
In section 3.1, I point out the three body decay distribution 
depends strongly  on the 
decay matrix element. This dependence may reduce the 
$m_{ll}$ end point resolution, while the whole shape 
of $m_{ll}$ distribution  could provide information on slepton masses and 
ino mixings.\cite{NY}
I also show a new analysis for the case where $\tilde{\chi}^0_2$ decays 
dominantly into 
$\tilde{l}l$.\cite{TKN} 
We point out that the peak position of $p_T$ asymmetry 
of the same flavor opposite sign (OS) lepton pairs 
in $m_{ll}\ll m_{ll}^{max}$ region would be  independent 
of $\tilde{\chi}^0_2$
momentum distribution, therefore may be used to constraint 
$m_{\tilde{\chi}^0_2}$, $m_{\tilde{\chi}^0_1}$ and 
$m_{\tilde{l}}$ directly such  as the end point of the  $m_{ll}$ distribution. 

In section 4, I discuss precision study at future LC's. Thanks to 
low backgrounds at polarized $e^+e^-$ collider, 
the machine is perfect to discover and study the
superparticles if they are in kinematical reach. Furthermore it offers
clean tests of relations of soft mass parameters and couplings. 
I discuss the radiative correction to the SUSY coupling relations
which can be probed precisely at LC. The measurement of the deviation
from the SUSY tree level relation offers a way to determine the squark mass
scale in the ``decoupling scenario''.\cite{nftetc}

\section{Supersymmetry and LHC}
Squarks ($\tilde{q}$) and gluinos ($\tilde{g}$) will be copiously 
produced at LHC, and they subsequently decay into 
charginos ($\tilde{\chi}^{\pm}_i$) or  
neutralinos ($\tilde{\chi}^{0}_i$).  
They could further decay into  sleptons ($\tilde{l}$). 
The signal of the sparticle production 
will be leptons  and/or jets with missing $p_T$ if LSP 
is stable\footnote{I will concentrate on the MSUGRA  
motivated scenario  where $\tilde{\chi}^0_1$ 
is the LSP.}.  Various study indicates that LHC will find the 
excess of the sparticle signal if $m_{\tilde{q}},
m_{\tilde{g}}< 2$ TeV in MSUGRA scenarios.  

The question is then if we could understand the nature of 
sparticles in detail.  MSSM  contains many parameters, on the other 
hand, the observed signal distributions are sum of 
products of production cross sections, branching ratios, 
and acceptances. The substantial complexities may  prevent 
simple and  model independent interpretations.

However some kinematical quantity can be extracted model independently
by investigating some characteristic decay distributions. One of
impressive examples is the case studied for Snowmass '96, so called
``LHC point 3''.  It is a case that the production of gluino followed
by $\tilde{g}\rightarrow\tilde{b} b$, $\tilde{b}
\rightarrow b \tilde{\chi}^0_2$  occurs with 
substantial branching fraction. 
The leptonic decay of the second lightest neutralino 
$\tilde{\chi}^0_2\rightarrow \tilde{\chi}^0_1 l^+l^-$ occurs 
with branching fraction of 16\%.  The number of  
$b\bar b b\bar b l^+l^- + 2$ jet events  then would be 
around 2.3 M for one year low luminosity run 
 with S/N ratio about 10:1; This is  
substantially  larger production ratio compared to typical s-channel 
sparticle production at LC.  The end point of $m_{ll}$ distribution
of  OS dileptons  
would be identified as $m_{\tilde{\chi}^0_2}$ $-m_{\tilde{\chi}^0_1}$.
The end point could  be  measured within 50 MeV error.

For point 3, $O(10^5)$ events near the $m_{ll}$ end point could be
selected for further analysis.  In the limit where $m_{ll}\sim
m_{ll}^{\rm max}$, $\tchi^0_1$ is stopped in the rest frame of
$\chi^0_2$, therefore $\vec\beta_{\tchi^0_2}\propto
\vec\beta_{\tchi^0_1}$ in the laboratory frame. Assuming further an
approximate MSUGRA relation $m_{\tilde{\chi}^0_2} $ $= 2
m_{\tilde{\chi}^0_2}$, one would reconstruct $m_{\tilde{b}}$ and
$m_{\tilde{g}}$ through the invariant mass distribution of all
possible combination of bottom jets and $\tilde{\chi}^0_2$
momentum. This leads to the resolution of MSUGRA parameters $m_0= 200
^{+13}_{-8}$ GeV and $M= 100 \pm 0.7$ GeV.

The above analysis showxs that the event distribution (which in
principle depends on hundreds of parameters of MSSM model) could be
factrized into a few distributions which sensitively reflects a few
parameters of the model.  The rest of the distributions will be
understood better with the constraints.  To this end, we may be able
to provide enough cross checks between events and theoretical
calculations (or MC simulations), so that we would be able to use
event rates and whole distributions to determine model parameters
precisely, or even reject some SUSY breaking scenarios.

\section{Neutralino decay into leptons}

In the previous section, we find 
the invariant mass distribution of 
OS lepton pairs from $\tilde{\chi}^0_2$ decay is the important 
part of the analysis. 
In this section we concentrate on some new aspects on the 
nature of the decay distribution and discuss the constraints to 
MSSM parameters that would be obtained from the distribution measurement.

\subsection{Three body decay into $\tilde{\chi}^0_1 l^-l^+$ 
and the decay matrix element}

Three body decays of $\tchi^0_2$ are dominant as long as two body
decays such as $\tchi^0_2\rightarrow\tchi^0_1Z^0$, $\tchi^0_1 H$,
$\tchi^0_2\rightarrow\tilde{l}l$ are not open. The branching ratio of
the three body leptonic decay of the second lightest neutralino,
$\tchi^0_2\rightarrow\tchi^0_1l^+l^-$, is known to be very sensitive
to the values of the underlying MSSM parameters. The dependence is
enhanced by the negative interference between the decay amplitude from
$Z^0$ exchange and that from slepton exchange. In this subsection we
show that the effect of the interference appears not only in the
branching ratios, but also in the decay distributions, such as the
distribution of the invariant mass $m_{ll}$ of the lepton pairs.

The partial decay width is given by 
\begin{equation}\label{e5}
\frac{d\Gamma}{dx\,dy}(\tchi^0_A\rightarrow\tchi^0_B\bar{f}f)=
\frac{N_C}{256 \pi^3} m_{\tchi^0_A} \vert{\cal M}\vert^2(x,y,
z=1+ r_{\tchi_B}^2-x-y).
\end{equation}
The range of ($x$, $y$) is given by the conditions
\begin{eqnarray}\label{e6}
&z (xy-r_{\tilde{\chi}_B}^2) \ge 0,&
\nonumber\\
&r_{\tchi_B}^2 \le x \le 1,&
\nonumber\\
&r_{\tchi_B}^2 \le y \le 1,&
\nonumber\\
&x+y+z=1+ r_{\tchi_B}^2,&
\label{psd}
\end{eqnarray}
when $m_f=0$. 

In the phase space of the decay $\tchi^0_2\rightarrow\tchi^0_1l^+l^-$,
the $Z^0$ exchange amplitude and $\tilde{l}$ exchange amplitude behave
differently. When the $Z^0$ contribution dominates, distributions are
enhanced in the region of large $m^2_{ll}=zm_{\tchi^0_2}^2$.  In
contrast, when the $\tilde{l}$ exchange contribution dominates,
distributions are enhanced in regions with large $x$ and/or large $y$,
therefore in small $m_{ll}$ and large $\vert E^{\rm rest}_{l^-}-E^{\rm
rest}_{l^+}\vert$ region. Note that $E^{\rm rest}_{l^+}= $ $
(1-x){m_{\tchi^0_2}}/2$ and $E^{\rm rest}_{l^-}= $ $
(1-y){m_{\tchi^0_2}}/2$ are lepton energies in the $\tchi^0_2$ rest
frame.

We consider the case where $2M_1\sim M_2 \ll |\mu|$, a typical case 
in MSUGRA model. In this case, $\tchi^0_2$ is 
Wino-like and $\tchi^0_1$ is Bino-like. 
An interesting property in this case is that the $Z^0$ and $\tilde{l}$ 
amplitudes could be of comparable size in some region of 
phase space. Furthermore, their interference is generally 
destructive for leptonic decays. These effects cause 
complicated situations, which we discuss below. 

For illustration, 
we use two sets of parameters for the neutralino sector, (A) and (B), 
shown in Table \ref{nsets}. 
\begin{table}[htb]
\begin{center}
\begin{tabular}{|c|c|c|c|c|} \hline
$\Biggl.$set & $M_1$ & $M_2$ & $\mu$ & $\tan\beta$ \\[2mm]\hline
$\Biggl.$(A)   & 70    & 140   & --300 & 4           \\[1mm]\hline
$\Biggl.$(B)   & 77.6  & 165   & 286   & 4           \\[1mm]\hline
\end{tabular}
\end{center}
\caption{\footnotesize Parameter sets for neutralinos. 
All entries with mass units are in GeV.}
\label{nsets}
\end{table}
These values are fixed to give the same masses for three inos, 
($m_{\tchi^0_1}$, $m_{\tchi^0_2}$, $m_{\tchi^+_2}$) $=$ 
(71.4, 140.1, 320.6) GeV. For calculating the branching ratios, 
we take generation-independent slepton masses and 
a universal soft SUSY breaking squark mass $m_{\tilde{Q}}=500$ GeV.

In Fig. 1a, we show the $m_{ll}$ distribution of the decay
$\tchi^0_2\rightarrow l^+ l^- \tchi^0_1$ for parameter set (A) and
varying $m_{\tilde{l}}$ from 170 GeV to 500 GeV. Because
$m_{\tchi^0_2}$ and $m_{\tchi^0_1}$ are fixed, the end points of the
distributions $m_{ll}^{\rm max}=68.7$ GeV are same for each curve,
while the shape of the distribution changes drastically with slepton
mass. For a slepton mass of 170 GeV (thick solid line), the decay
proceeds dominantly through slepton exchanges, therefore the $m_{ll}$
distribution is suppressed near $m_{ll}^{\rm max}$. On the other hand,
once slepton exchange is suppressed by its mass, $Z^0$ exchange
dominates and the distribution peaks sharply near $m_{ll}^{\rm max}$.

In Fig. 1b, we show an example for $\mu>0$, parameter set (B).
The dependence on the slepton mass is different from the previous 
case. As $m_{\tilde{l}}$ increases from 170 GeV, $m_{ll}$ 
distribution becomes softer. For 
$m_{\tilde{l}}> 250$ GeV, a second peak appears 
due to strong cancellation of $Z^0$ exchange 
and slepton exchange contributions for a certain value of 
$m_{ll}$.  At the same time, the branching ratio reaches its minimum 
at $m_{\tilde l}\sim 300$ GeV, much less than 1\%. 

\begin{figure}[htbp]
\begin{center}
\includegraphics[width=5.5cm,angle=-90]{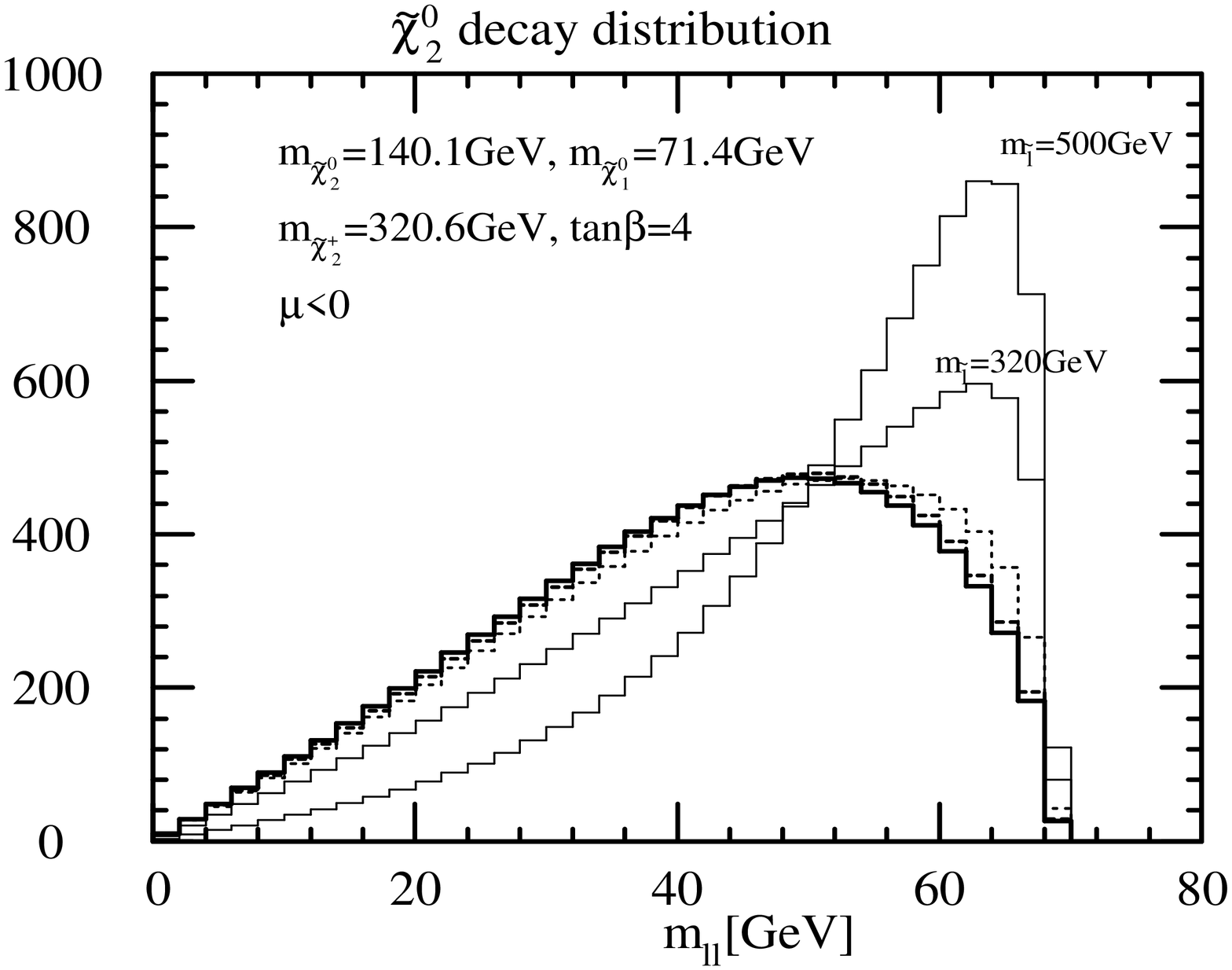}
\hskip 2cm 
\includegraphics[width=5.5cm,angle=-90]{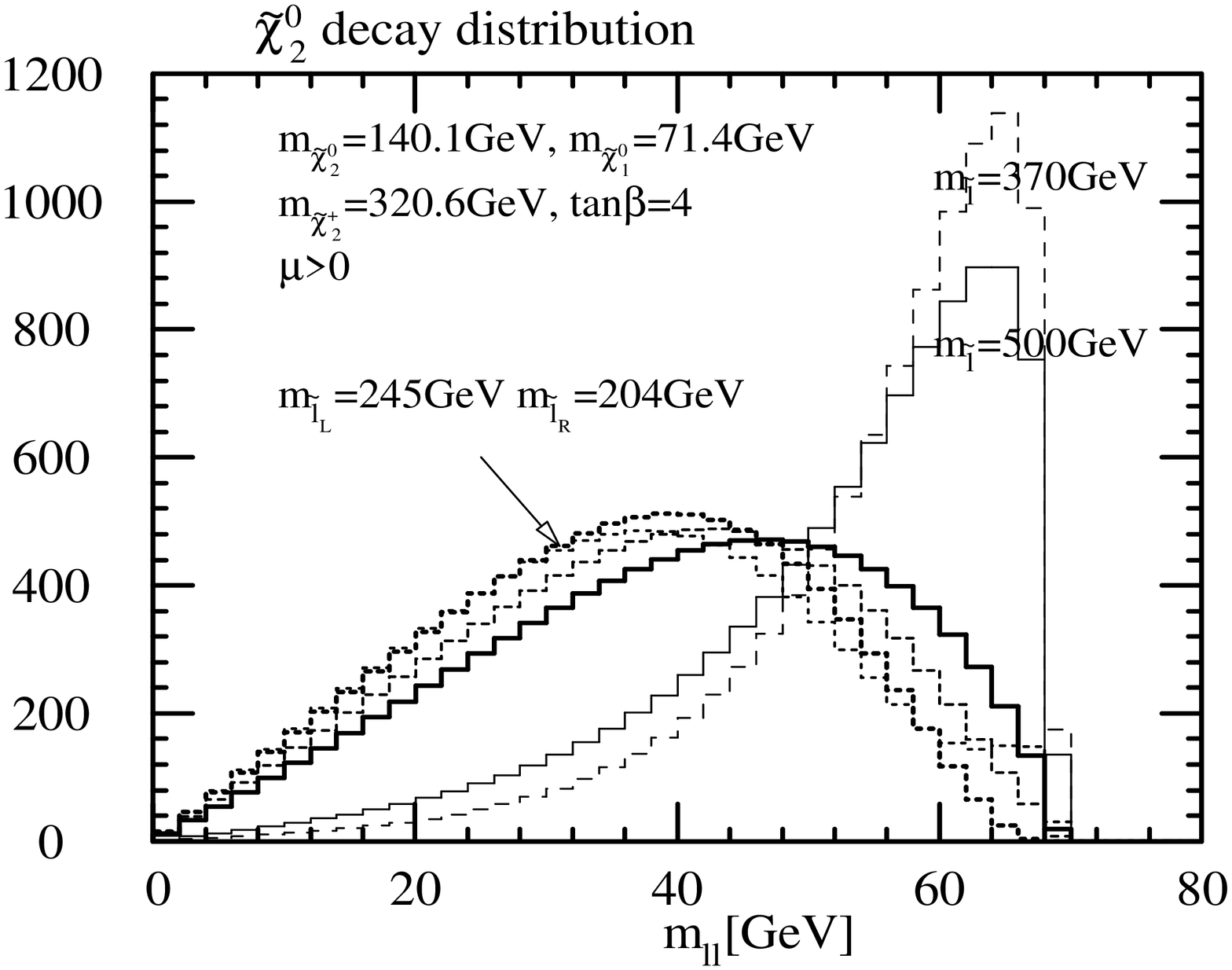}
\end{center}
\hskip 3.5cm 
a)  
\hskip 6cm 
b)
\caption{\footnotesize  
$m_{ll}$ distribution of $\tilde{\chi}^0_2\rightarrow $
$\tilde{\chi}^0_1 ll$ decay for different
$m_{\tilde{l}}$. a) $\mu <0$, and  b) $\mu >0$. 
$m_{\tilde{\chi}^0_2}-m_{\tilde{\chi}^0_1}$ is 
fixed throughout the plots. }
\label{fig1}
\end{figure}

Notably, one can find slepton masses 
where a complete cancellation occurs very 
close to the end point $m^{\rm max}_{ll}$ of the $m_{ll}$ distribution. 
The thick dashed line shows distribution for 
$m_{\tilde{l}_L}=245$ GeV and $m_{\tilde{l}_R}=204$ GeV. 
Events near the end point 
($m^{\rm max}_{ll}-m_{ll}< 4$ GeV) becomes too few, 
and it is very hard to observe the real end point 
for this case.

As it has discussed already, the lepton invariant mass distribution is
an important tool for studying MSSM at hadron colliders. In previous
studies, the end point of $m_{ll}$ distribution is treated as
ambiguous measurement of $m_{\tilde{\chi}^0_2}- m_{\tilde{\chi}^0_1}$,
and the end point samples are used for further analysis.  The slepton
mass dependence of $\tchi^0_2$ decay distribution shown in Fig. 1
suggests that not only the end point of the distributions but also the
distributions themselves contain information about the underlying
parameters such as $m_{\tilde{l}}$. The negative side of this is that
the fitted end point may depend on the assumed values of these
parameters, introducing additional systematic errors to the fit. For
an extreme case shown in Fig. 2, the observed end point of the lepton
invariant mass distribution does {\it not} coincide with
$m_{\tchi^0_2}-m_{\tchi^0_1}$. Note that realistic simulations
including the parameter dependence of the decay distribution were not
available for hadron colliders until recently. The most recent ISAJET
release ($>$ISAJET 7.43) allows to simulate the effect of exact matrix
elements for all three body decay distributions.

The $m_{ll}$ distribution may be used to extract the underlying MSSM
parameters. The distribution depends strongly on $m_{\tilde{l}}$, and
also on $\tchi_2^0 l\tilde{l}$ and $\tchi^0_2\tchi^0_1Z^0$ couplings.
The $Z^0$ coupling is proportional to the Higgsino components of
$\tchi^0_2$ and $\tchi^0_1$. Because we take these neutralinos to be
gaugino-like, their Higgsino components depends on $\tan\beta$ and the
Higgsino mass parameter $\mu$, and $|\mu|$ is roughly equal to
$m_{\tchi^+_2}$. Therefore the $m_{ll}$ distribution gives at least
one constraint on $\mu$, $\tan\beta$ and $m_{\tilde{l}}$ in addition
to the well known constraint on $m_{\tchi^0_2}-m_{\tchi^0_1}$.

We estimate the sensitivity, 
assuming that backgrounds can be neglected or subtracted, 
and dependence of acceptance on $m_{ll}$ can be corrected.
We define the sensitivity 
function ${\cal S}$ as follows: 

\begin{equation}\label{e9}
{\cal S}=\sqrt{\sum_i 
\left(n_i^{\rm fit}-n_i^{\rm input}\right)^2/n_i^{\rm input} }\, .
\end{equation}
Here $n_i^{\rm fit}$ ($n_i^{\rm input}$) is the number 
of events in the $i$-th bin of the $m_{ll}$ distribution 
for the MSSM parameters 
$(M_1, M_2, \mu, \tan\beta, m_{\tilde l_{L, R}})
\vert_{\rm fit(input)}$. We normalize $\sum_i n_i^{\rm fit}$ and  
$\sum_i n_i^{\rm input}$ to some number $N$.   
${\cal S}$ gives the deviation of 
the input distribution $n_i^{\rm input}$  from the distribution 
for the fit ($n_i^{\rm fit}$) in units of standard deviations.
We take 
$N=2500$ and an $m_{ll}$ bin size of 2 GeV.

In Fig. 2, we show contours of constant ${\cal S}=1, 2, 3, 4$
(corresponding to 1$\sigma$, 2$\sigma$, 3$\sigma$, 4$\sigma$ for
N=2500) in the ($\tan\beta^{\rm fit}$, $m_{\tilde{l}}^{\rm fit }$)
plane. For the solid lines, we take parameter set (A) and
$m_{\tilde{l}}=$ 250 GeV as input parameters, while for fitting
parameters we vary $\tan\beta$ and $m_{\tilde{l}_L}=m_{\tilde{l}_R}$,
fixing ($M_1^{\rm fit}$, $M_2^{\rm fit}$, $\mu^{\rm fit}$) to
reproduce the input values of ($m_{\tchi^0_1}$, $m_{\tchi^0_2}$,
$m_{\tchi^+_2}$).  The resulting contours (solid lines) correspond to
the sensitivity of the $m_{ll}$ distribution to $m_{\tilde{l}}$ and
$\tan\beta$ when the three ino masses are known.

\begin{figure}[htbp]
\begin{center}
\includegraphics[width=6cm,angle=-90]{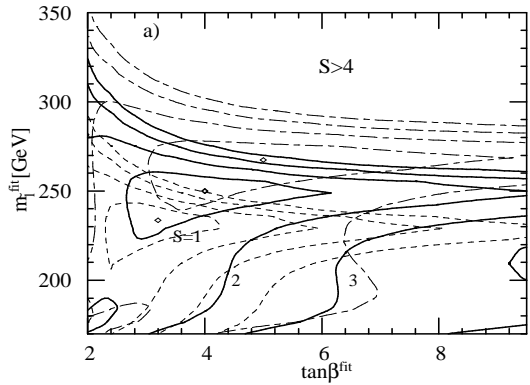}
\end{center}
\label{fig2}
\caption{\footnotesize Constraint on $m_{\tilde{l}}$ and $\tan\beta$  from the
$m_{ll}$ distribution.
Input parameters are  set (A) with 
$m_{\tilde{l}}=250$ GeV. For solid lines, 
we fix $m_{\tchi^0_2}$, $m_{\tchi^0_1}$ and  $m_{\tchi^+_2}$ equal to 
those for parameter set (A), while 
$\tan\beta^{\rm fit}$ and $m_{\tilde{l}}^{\rm fit}$ are 
varied to see the sensitivity of the $\tchi^0_2$ 
decay distribution to these parameters.
For the dot-dashed (dashed) lines, $m_{\tchi^+_2}^{\rm fit}
=m_{\tchi^+_2}^{\rm input}+(-) 30$ GeV. 
 }
\end{figure}

In the figure, a strong upper bound on the slepton masses emerges,
$m_{\tilde{l}}<260 $ GeV is obtained if ${\cal S}<1 $ is
required. This is consistent with the large change of the
distribution between $m_{\tilde{l}}=270$ GeV and $m_{\tilde{l}}=500$
GeV found in Fig. 1. 
The $m_{ll}$ distribution also constrains $\tan\beta$ mildly. 
The constraint is not very strong due to our 
choice of parameters $|\mu|\gg M_2$; gaugino-Higgsino  
mixing is suppressed in this case.

\subsection{The two body decay into $\tilde{l} l $ 
and lepton $p_T$ asymmetry}

We now discuss the case where $m_{\tilde{\chi}^0_2}>m_{\tilde{l}}$.
The decay could proceed through two body decays
\begin{eqnarray}
\tilde{\chi}^0_2& \rightarrow & \tilde{l}^+l^-, \tilde{l}^-l^+,  \\ 
\tilde{l}&\rightarrow &\tilde{\chi}^0_1 l.
\end{eqnarray}
Because of the phase space factor, the decay  could dominate over the 
three body decay $\tilde{\chi}^0_2\rightarrow \tchi^0_1 f\bar{f}$.
 
Note that the decay distribution is now completely fixed by 
the two body kinematics.  The $m_{ll}$ distribution of the 
two leptons from the $\tilde{\chi}^0_2$  cascade decay is 
\begin{equation}
\frac{1}{\Gamma}\frac{d\Gamma}{d m_{ll}}=  \frac{2 m_{ll}}
{(m^{\rm max}_{ll})^2}.
\end{equation}
Here the end point of the $m_{ll}$ distribution, $m_{ll}^{\rm max}$, 
is expressed as 
\begin{equation}
m_{ll}^{\rm max}=\frac{\sqrt{(m^2_{\tilde{\chi}^0_2}-m^2_{\tilde{l}})
(m^2_{\tilde{l}}-m^2_{\tilde{\chi}^0_1})}}{m_{\tilde{l}}}\ \ . 
\end{equation}

In addition to that, it has been known that the the lepton $p_T$
asymmetry $A_T (\equiv p_{T2}/p_{T1})$ distribution (where $p_{T1}>
p_{T2}$) is sensitive to slepton masses.\cite{H,IK} The $A_T$ can
distribute off from 1 when it originates from $\tilde{\chi}^0_2$ and
sometimes strongly peaks. (On the other hand, $A_T$ distribution of
the three body decay peaks at 1.)  The asymmetry comes from the
monochromatic nature of lepton energy from the $\tilde{\chi}^0_2$
decay in the rest frame. For example, when the mass difference between
$\tilde{\chi}^0_2$ and $\tilde{l}$ is small, the lepton energy from
the $\tilde{\chi}^0_2$ decay is substantially smaller than that from
$\tilde{l}$ decay. The nature of the lepton $p_T$ asymmetry is then
qualitatively understand, because the lepton with high (low) energy in
the $\tilde{\chi}^0_2$ rest frame has better chance to get high (low)
$p_T$ in the laboratory frame.

The purpose of this subsection is to improve this qualitative nature 
to quantitative one.  We note that two leptons 
go exactly the same direction if $m_{ll}= 0$. 
In the limit, the ratio of the energies of the lepton antilepton 
pair  is unchanged 
even if $\tilde{\chi}^0_2$ is boosted; $A_T$ at $m_{ll}=0$, 
$A^0_T$, would be estimated by the lepton energy ratio $A_E(\equiv 
E_{l2}/E_{l1})$ at 
the  $\tilde{\chi}^0_2$ rest frame at $m_{ll}=0$, $A_E^0$,  as, 
\begin{eqnarray} 
A^0_T \sim A^0_E \equiv E_{l2}/E_{l1}\vert_{m_{ll}\sim 0} &=& 
\frac
{m^2_{\tilde{l}}- m^2_{\tilde{\chi}^0_1} }
{ m^2_{\tilde{\chi}^0_2}-m^2_{\tilde l}},\ \ 
({\rm for\ } 
m^2_{\tilde{l}}- m^2_{\tilde{\chi}^0_1}< 
m^2_{\tilde{\chi}^0_2}-m^2_{\tilde l} )
\cr 
\ \ \ {\rm or} 
& & \frac
{ m^2_{\tilde{\chi}^0_2}-m^2_{\tilde l}} 
{m^2_{\tilde{l}}- m^2_{\tilde{\chi}^0_1} }
\ \ \   ({\rm for\ } 
m^2_{\tilde{l}}- m^2_{\tilde{\chi}^0_1}> 
m^2_{\tilde{\chi}^0_2}-m^2_{\tilde l} ). 
\cr &&
\end{eqnarray}

Even though $m_{ll}\neq 0$, we see that distribution peaks at the same
value; $A^{\rm peak}_T= A^0_E$ holds approximately. In Fig 3, we show
$A_T$ distributions with various invariant mass cuts. Here we take the
universal scaler mass $m = 100$ GeV, the universal gaugino mass $M =
150$ GeV, $\tan\beta = 2$, $A = 0$, and $\mu < 0$, when
$m_{\tilde{e}_R }= 120.68 $ GeV, $m_{\tilde{\chi}^0_1}= 65.15$ GeV and
$m_{\tilde{\chi}^0_2}= 135.49$ GeV.  The peak structure is quite
prominent even without $m_{ll}$ cuts.  The edge of the $m_{ll}$ decay
distribution is 52 GeV.\footnote{The MSSM parameters and cuts are same
to that is taken by Iashvili and Khar-chilava\cite{IK}. We use
ISAJET7.44 and ATLFAST2.21 for simulations.}

\begin{figure}[htbp]
\begin{center}
\includegraphics[width=3.5cm]{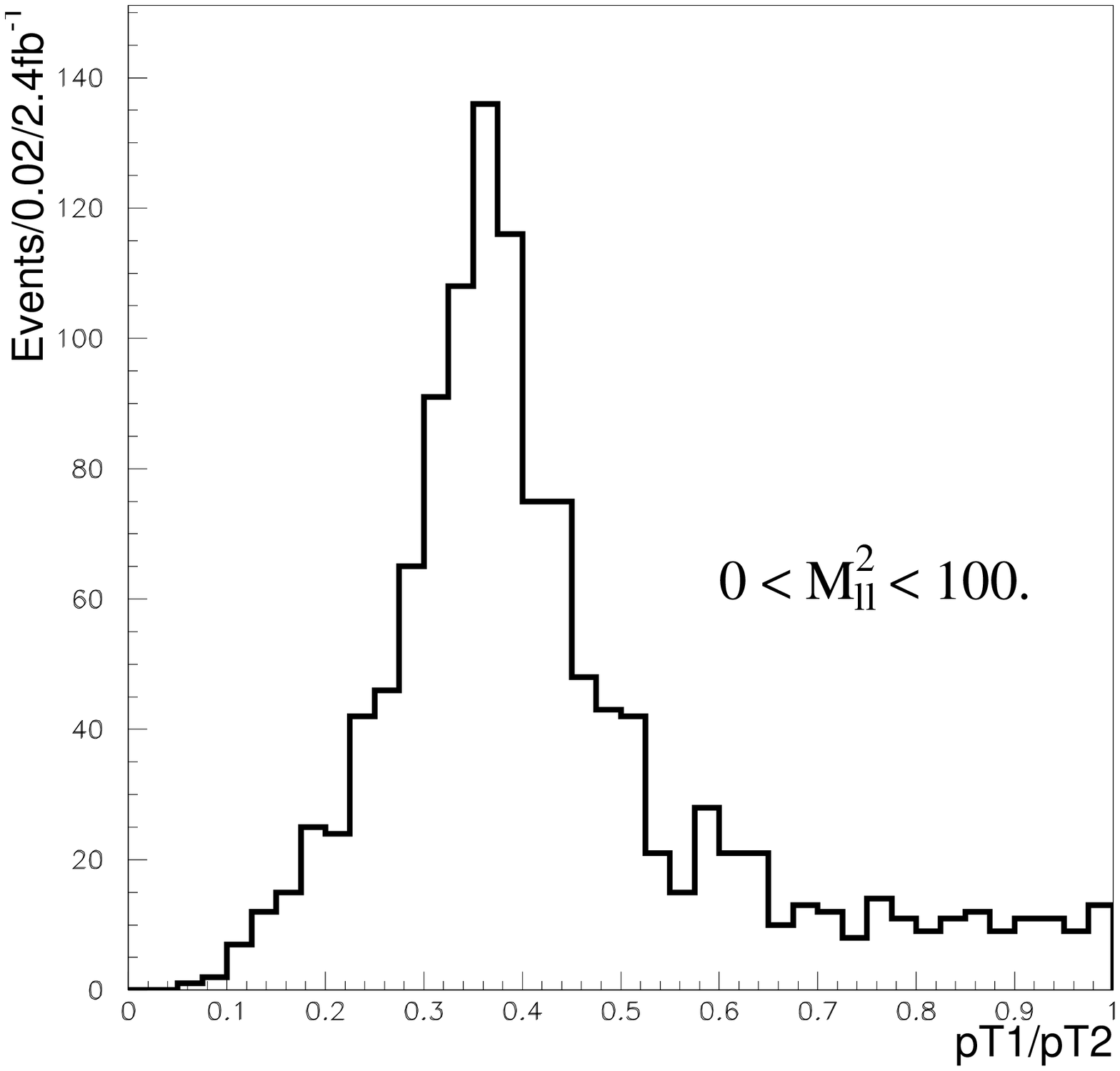}
\hskip 0.3cm 
\includegraphics[width=3.5cm]{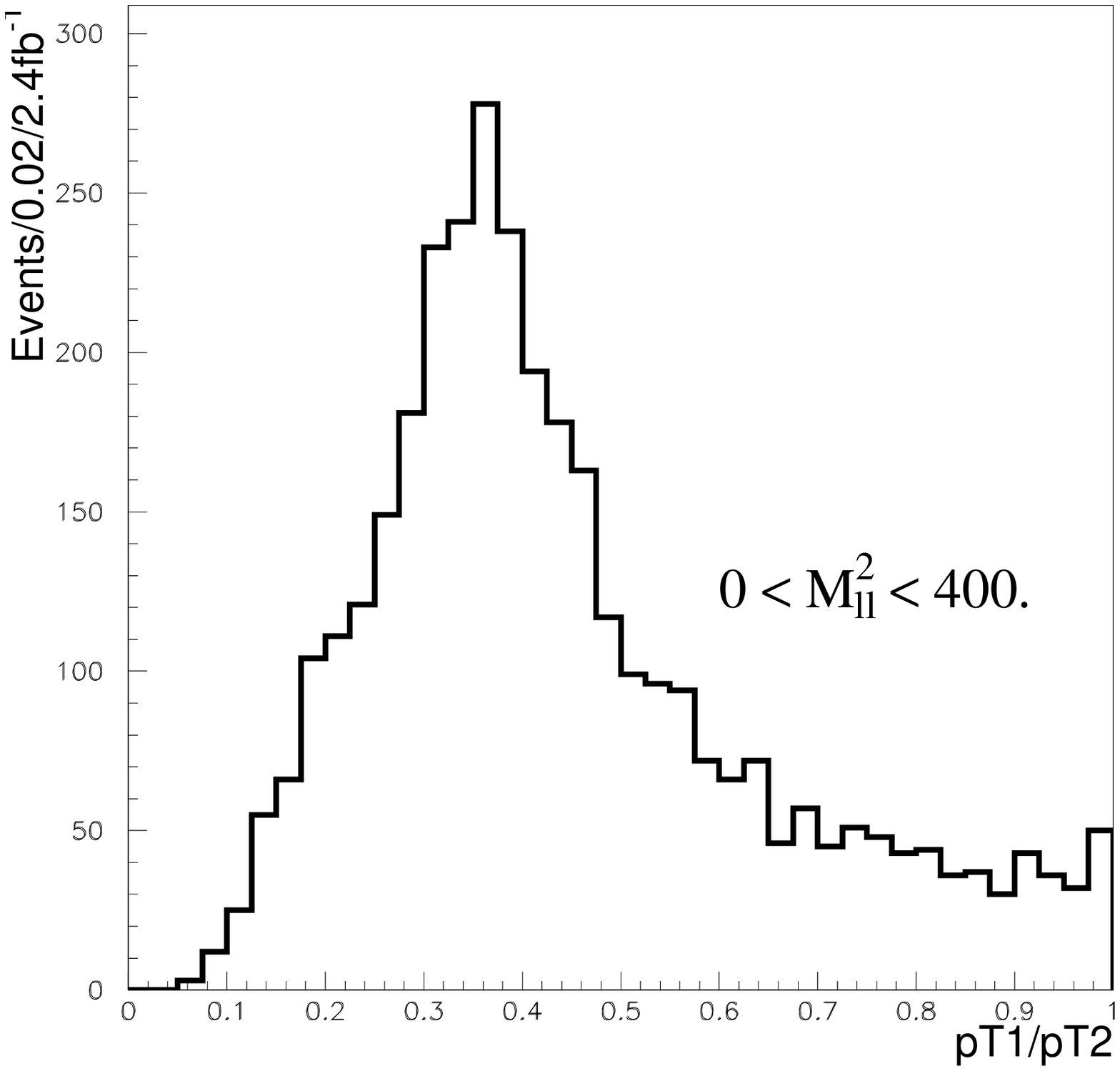}
\hskip 0.3cm 
\includegraphics[width=3.5cm]{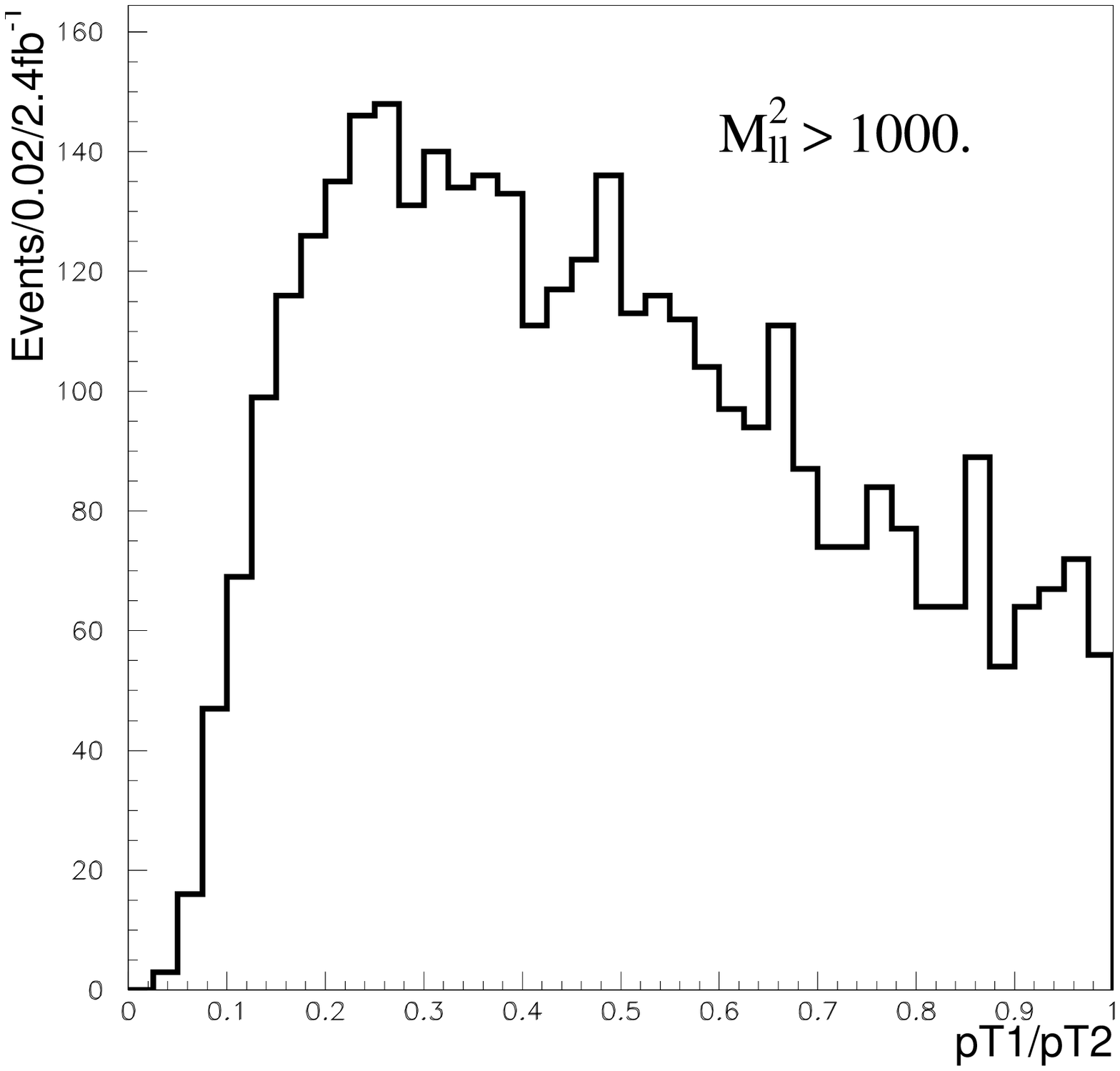}
\end{center}
\hskip 2cm a) \hskip 4cm b) \hskip 4cm c) 
\caption{
\footnotesize $A_T\equiv p_{T2}/p_{T1}$ distribution with different 
$m_{ll}$ cut. a)$m^2_{ll}<100$ (GeV)$^2$,
b) $m^2_{ll}<400$ (GeV)$^2$, c) $m^2_{ll}>1000$ (GeV)$^2$ 
}
\label{fig3}
\end{figure}

We can see that the peak solely comes from $m_{ll}\ll m^{max}_{ll}$
events. When we compare the distribution of the events with
$m^2_{ll}<100$ (GeV)$^2$ (Fig.3 a) and $m^2_{ll}<400$ (GeV)$^2$ (Fig.3
b), we found that the event distribution is more sharply peaked at the
value close to $A_E^0 = 0.368$ as $m_{ll}$ cut decreases.  The result
of non-symmetric Gaussian fit to the events near the peak for
$\int{\cal L} dt = 2.4 fb^{-1}$ is summarized in the table 1. The
sample for $m_{ll}<10$ GeV is in perfect agreement with the expected
value $A_E$.

With sufficient statistics one may be able to
measure the $m_{ll}$ dependence of $A_T$ in  $m_{ll}\ll m_{ll}^{max}$ 
region. The lepton energy ratio for generic $m_{ll}$ would be given as
\begin{eqnarray}\label{gmll}
\frac{E_{l2}}{E_{l1}} &=& 
\frac{m^2_{\tilde{\chi}^0_2} -m^2_{\tilde{l}}}
{m^2_{\tilde{l}} -m^2_{\tilde{\chi}^0_1}+ 2 m^2_{ll}   }
\ \ ({\rm for }\  m^2_{\tilde{l}} -m^2_{\tilde{\chi}^0_1}+ m^2_{ll}  >
m^2_{\tilde{\chi}^0_2} -m^2_{\tilde{l}})
\cr
&{\rm or} &\frac{m^2_{\tilde{l}} -m^2_{\tilde{\chi}^0_1}+ m^2_{ll}   }
{m^2_{\tilde{\chi}^0_2} -m^2_{\tilde{l}}} 
\ \ ({\rm for }\  m^2_{\tilde{l}} -m^2_{\tilde{\chi}^0_1}+ m^2_{ll}  <
m^2_{\tilde{\chi}^0_2} -m^2_{\tilde{l}}).
\end{eqnarray}
According to Eq.(\ref{gmll}), the measurement of 
$m_{ll}$ dependence of the peak 
position  could correspond to the
measurement of $m^2_{\tilde{l}}-m^2_{\tilde{\chi}^0_1}$.

In Table 2, the fitted peak value reduces as $m_{ll}$ cut
increases. This could be due to the reduction of average
$E_{l1}/E_{l2}$ in the $\tilde{\chi}^0_2$ rest frame. $A_E = 0.354$ at
$m_{ll}=20$ GeV for the parameters we take.  The average $A_E$ between
$m_{ll}=0$ and 20 GeV agrees with the peak value of $A_T$
distribution, though the statistics of this simulation is not
sufficient to claim the deviation of the peaks between $m^2_{ll}<100$
(GeV)$^2$ and $m^2_{ll}<400$ (GeV)$^2$ samples. On the other hand,
$A_E= 0.2915$ at $m_{ll} = m_{ll}^{\rm max}= 52$ GeV. The average
$A_E$ between $m_{ll}=0$ to $m_{ll}= m_{ll}^{\rm max}$ is too small
compared to the $A^{\rm peak}_T$ without $m_{ll}$ cut. This is
consistent with the fact that no peak structure is observed for
$m^2_{ll}> 1000 ({\rm GeV})^2$ (Fig. 3c).
\vskip 0.5cm

\begin{table}[htb]
\begin{center}
\begin{tabular}[b]{|c||c|c|c|}
\hline
& $m^2_{ll}<100$ (GeV)$^2$   &  $m_{ll} <400$ (GeV)$^2$ & no $m_{ll}$  cut 
\\
&  $0.30<A_T< 0.40$ & $0.20 <A_T< 0.40$ & 
$0.15< A_T <0.40$
\\\hline\hline
peak value & 0.362 & 0.352 & 0.349 
\\\hline
Error & $0.660\times 10^{-2}$ & $0.637\times 10^{-2}$ & $0.685\times 10^{-2}$ 
\\\hline
\end{tabular}
\end{center}
\caption{\footnotesize Peak position and its error in $A_T$ distribution}
\label{fit}
\end{table}

To see the importance of $A_T$ measurement, we first show the expected
constraint on $m_{\tilde{l}}$ and $m_{\tilde{\chi}^0_1}$ when
$m_{\tilde{\chi}^0_2}$ is fixed, provided that $A_T$ and $m_{ll}^{\rm
max}$ is measured with the error of 0.07 and 0.5 GeV respectively
(Fig.4). The error on $m_{\tilde{l}}$ and $m_{\tilde{\chi}^0_1}$ could
be of the order of 1\%, consistent with the previous
fits\cite{IK}. Note they did not identify the origin of the peak
structure and used {\it whole} $A_T$ distribution for the fit. The
used distribution may depend on parent squark and gluino masses, while
our fit relays solely on the peak position, or only on
$m_{\tilde{\chi}^0_1}$, $m_{\tilde{\chi}^0_2}$ and $m_{\tilde{l}}$.

\begin{figure}[htbp]
\begin{center}
\includegraphics[width=7cm,angle=-90]{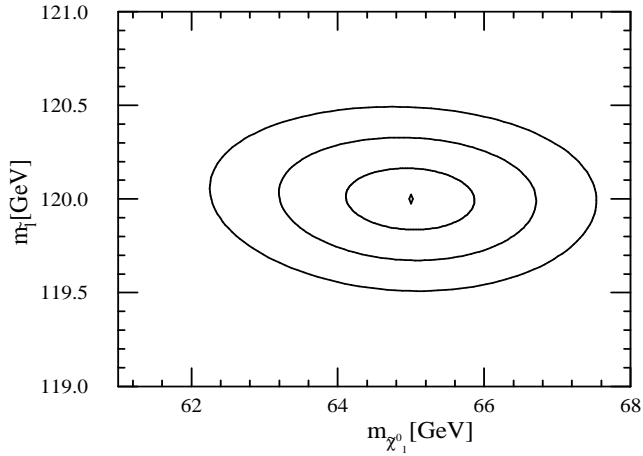}
\end{center}
\caption{\footnotesize Contours of constant $\Delta\chi^2=1,4,9$ 
when $\delta A^{\rm peak}_T<0.07$ and $\delta m_{ll}= 0.5$ GeV. 
$m_{\tilde\chi^0_2}$ is fixed for this fit. }
\label{fig4}
\end{figure}

One may also observe the end point of $m_{ll}$ distribution of the
three body decay in addition to the edge of the $m_{ll}$ distribution
originated from two body cascade decay through slepton. This is
because right handed slepton coupling to wino and higgsino is
zero. The measurements of $m_{ll}^{\rm max}$ (2 body), $A_T^0$, and
$m_{ll}^{\rm max}$ (3 body) $\equiv$
$m_{\tilde{\chi}^0_2}-m_{\tilde{\chi}^0_1}$ are potentially sufficient
to determine all sparticle masses involved in the $\tilde{\chi}^0_2$
cascade decay. Assuming a rather optimistic error on $m_{ll}^{\rm
max}({\rm 3\ body})$, $\delta m_{ll}^{\rm max}({\rm 3\ body})= 1$ GeV,
$m_{\tilde{\chi}^0_1}$, $m_{\tilde{\chi}^0_2}$, and $m_{\tilde{l}}$
are constrained within $\sim \pm 8$ GeV, without assuming any relation
between ino and slepton masses. The errors are substantially larger
than those shown in Fig.4, due to the correlations between the
constraints.  On the other hand, $m_{\tilde e}/m_{\tilde \mu}$ ratio
would be constrained strongly.  Assuming $\delta A_T<0.07$, $\delta
m_{ee, \mu\mu}<0.5$ GeV, $\delta m_{ll}^{\rm max}({\rm 3 body})= 4$
GeV, we obtain $\delta(m_{\tilde e}/m_{\tilde \mu})=2.5 $ \% for
$\Delta \chi^2 <1$, and 7\% for $\Delta\chi^2<9$.
 
Note that the background in $m_{ll}\ll m_{ll}^{max}$ region must be
studied to claim the above measurement is possible. Backgrounds from
$t\bar{t}l\bar{l}$ could be important in low $m_{ll}$ region. Note
that the full amplitude level study of $W\gamma^*$ production has been
done for the background process of
$\tilde{\chi}^0_2$ $\tilde{\chi}^+_1\rightarrow 3l$, and large background
is found in $m_{ll}<10$ GeV region\cite{baer}.  However it is unlikely
that the background distribution has peak in $0\ll A_T$ region. The
peak of the signal distribution may be observed precisely on the top
of such backgrounds, especially when signal rate is high enough to
allow precision studies.

\section{Precision study at LC}
A TeV scale linear colliders could be a powerful discovery machine. In
the context of MSUGRA model, 1 TeV LC roughly corresponds to LHC in
its discovery potential. This is because that in the model with
universal scalar and gaugino mass at very high scale, relations
$m_{\tilde{Q}}\gg m_{\tilde{l}}$ and $m_{\tilde{g}}\gg m_{\tilde{W}}$,
$m_{\tilde{B}}$ are predicted naturally. In future LC's, search modes
are the production and the decay of sleptons, charginos, and
neutralinos.

We should also note that LC experiments cover the case where 
superparticles takes nasty patterns of mass spectrum. 
Because of the available high beam polarization, backgrounds 
from $W^\pm$ boson pair production can be highly suppressed. 
Notice also  that its effective $\sqrt{s}$ is monochromatic  
for $e^+e^-$ colliders, 
therefore $\tilde{l}$ and $\tilde{\chi}^+_1$ will be produced
subsequently from lighter to heavier, 
and we can measure production cross sections and the decay distributions 
systematically. 

Systematical studies of physics potential at LC when 
slepton, chargino and  neutralino are produced have 
been done in detail by several authors\cite{JLCETC}, 
and it has been 
shown that gaugino mass relations, slepton mass relation, and coupling 
relations can be confirmed with errors of $O(1\%)$.  

In this note, we concentrate on  measurements of the coupling 
relations imposed by supersymmetry, 
\begin{eqnarray}
g_{\tilde{B}\tilde{e}_Re_R } &=& \sqrt{2}g_B, \cr
g_{\tilde{W}\tilde{e}_Le_L } &=& \sqrt{2}g_W.
\end{eqnarray}
It has been argued that this coupling relation could be 
measured within  $O(1\%)$ accuracy or better by measuring sparticle 
production cross sections, angular distributions, and 
sparticle masses involved in the production process \cite{nftetc,NPY}. 

The measurement of the couplings is important because 
the equivalence of the gauge coupling and gaugino 
coupling is ultimate probe of the supersymmetry at 
low energy, although the partial discovery of 
sparticle of course suggests the existence. 
Another way to say, the existence of (large) SUSY breaking sector 
couples to (observed) sparticles  will appear as the 
deviation of the sparticle  coupling from those 
predicted by the tree level symmetry. 

Such corrections might come from the existence of squarks which is 
much heavier than sleptons, charginos or neutralino. 
Such scenarios, with relatively light third generation squarks  
are occasionally quoted as ``decoupling'' scenarios, and  
attractive because they are free from large flavor 
changing neutral currents. When such mass spectrum is realized, 
SUSY coupling relations do not hold in the effective theory below 
$m_{\tilde{q}}$, and the corrections to the couplings 
from the tree level predictions are expressed as follows; 
\begin{eqnarray}
\delta \left(\frac{g_{\tilde{B}\tilde{e}e}}{g^{SM}_{\tilde{B}}}\right) 
& = & 0.7\% \log_{10}
\left(\frac{m_{\tilde{q}}}{m_{\tilde{l}}}\right),\cr
\delta \left(\frac{g_{\tilde{W}\tilde{e}e}}{g^{SM}_{\tilde{W}}}\right)
 & = & 2\% \log_{10}
\left(\frac{m_{\tilde{q}}}{m_{\tilde{l}}}\right).
\end{eqnarray}
If $t$-channel exchange of sparticles dominates over $s$-channel 
exchange of gauge bosons, 
the cross section can be proportional to 
the 4th power of the coupling, and the correction to the cross section 
could be around 8\% when $m_{\tilde{q}}=10 m_{\tilde{l}}$ 
for sneutrino and wino productions.  
A specific example is considered for
$\tilde{\nu}\tilde{\nu}^*$ production with charginos lighter 
than $\tilde{\nu}$\cite{NPY}. The production 
is dominated by $t$ channel exchange of charginos, 
and involve the wino-sneutrino-electron coupling. 
For integrated luminosity around $100 fb^{-1}$, 
the accepted number of events consisted with $e^+e^-$ and some 
other jets or leptons activity exceeds more than $10^4$ events.   With 
suitable constraint to $\tan\beta$ and heavier ino masses, 
$\log_{10} ( m_{\tilde q }/m_{\tilde{l}})$ would be constrained 
within 0.09 (statistics) $\pm$ 0.08 (sneutrino mass error).
\footnote{
Here the error from sneutrino mass uncertainty is relatively 
large due to $\beta_{\tilde{\nu}}^3$ behavior of the cross section near 
the sneutrino production threshold. 
Chargino threshold production is proportional 
to $\beta_{\tilde{\chi}}$ and the error due to the mass uncertainty
might be controlled better.}

It would  not be very surprising if 
${\cal O}(10^6)$ sparticle events is accumulated in future, 
with sufficient 
understanding on underlying parameter of MSSM models. 
Note that different sparticles produced simultaneously,
and  the  proposed TESLA integrated luminosity is as large as  1 ab$^{-1}$= 1000fb$^{-1}$. Then 
Does this mean that the production cross sections 
are measured within  O(0.1\%) accuracy;
a few \% measurement of the first generation squark masses without 
producing them?

Apparently, measuring the  number of signal events is not equivalent to 
the measurement of the production cross section.  
Measured production cross sections suffer 
various uncertainties, which is schematically given as 
\begin{eqnarray}\label{e12}
\frac{\delta\sigma}{\sigma}&=&\frac{1}{\sqrt{N_{\rm accept}}}
\oplus \frac{\delta\sigma}{\delta M_i }\delta M_i \oplus
\frac{\delta\sigma}{\delta m_i }\delta m_i \oplus
...\cr  
& &\oplus {\rm luminosity\  error} \oplus 
{\rm energy\  resolution}\oplus{\rm QED,\  QCD\  corrections} ...\cr
&&
\end{eqnarray}
where $N_{\rm accept}$ may be expressed  as 
\begin{eqnarray}
N_{\rm accept}&=& Br ({\rm sparticle\ \rightarrow\  visible\ or\ clean 
\  mode}) \cr
&&\times 
({\rm acceptance}) 
\times (\int dt {\cal L} \sigma ).
\end{eqnarray}
The branching ratios  may be 
around $(50\%)^2$ and the acceptance  could be as high as  50 \%. 
In the right hand side of equation (\ref{e12}), the 
first line contains errors of underlying MSSM parameters 
that could be negligible in the limit 
of infinite statistics.
The second line contains machine dependent errors and 
potentially large and uncontrolled QED and QCD corrections. 
They must be very small if we want to extract $<1\%$ deviation of 
cross section, and would require the huge efforts. 

\section{Conclusion}

In this talk, I discussed a ``precision'' study of 
supersymmetry in future colliders, LHC and LC. 

The motivation of the precision study is to explore the origin of 
supersymmetry breaking and mechanisms to bring it 
to our sector. The signature must appear on the 
sparticle mass patterns, and would be studied in detail in 
LHC and LC. 

For LHC, charginos and neutralinos are produced as 
decay products of $\tilde{g}$ and $\tilde{q}$, and 
the nature of weak interacting sparticles will be studied. 
In this talk, I discussed 
the decay of the second lightest neutralino. 
The leptonic decays of the second lightest neutralino 
$\tilde{\chi}^0_2$ could be studied even though one does not know the parent 
neutralino momentum. 
The $m_{ll}$  distribution of  the three body decay 
is sensitive to neutralino 
mixing and slepton masses. If systematical errors can be 
controlled, one may be able to constrain slepton masses. 
When the two body decay $\tilde{\chi}^0_2\rightarrow \tilde{l}$ is 
open, 
one would measure the peak of lepton $p_T$ asymmetry in 
$m_{ll}\ll m_{ll}^{\rm max}$. 
in addition to the edge of  $m_{ll}$ distribution of the two body decay 
and occasionally the $m_{ll}$ end point of 
the three body . The information constrain the parent and 
daughter neutralino and slepton masses 
rather stringently in model independent manner. 

In LC, not only masses of sparticles, 
but  production cross sections and sparticle decay 
distributions 
will be measured precisely. Underlying MSSM parameters, such as 
sparticle soft mass parameters, $\tan\beta$, gaugino-sfermion-fermion 
coupling would  be measured within  precision of $O(1\%)$ or less. 
In this talk, we discuss the determination of 
squark mass scale  in the ``decoupling scenario'' where $\tilde{q}$ is 
much heavier than $\tilde{W}, \tilde{B}$, and $\tilde{l}$.
The squark mass maybe constrained within O(10\%) through the measurement 
of the deviation of $\tilde{W}\tilde{l} l $ coupling 
from its tree level value, $g^{SM}_2$.


\end{document}